\documentclass{jltp}

\usepackage{graphicx}

\begin{document}

\title{Evaluation of specific heat for superfluid helium between 0 - 2.1 K based on nonlinear theory}

\author{Shosuke Sasaki}

\address{Shizuoka Institute of Science and Technology, Fukuroi, Shizuoka, 437-8555, Japan}

\maketitle

\begin{abstract}
The specific heat of liquid helium was calculated theoretically in the Landau theory \cite {Landau}. The results deviate from experimental data in the temperature region of 1.3 - 2.1 K. Many theorists subsequently improved the results of the Landau theory by applying temperature dependence of the elementary excitation energy \cite {Bendt}, \cite{Brooks}. As well known, many-body system has a total energy of Galilean covariant form. Therefore, the total energy of liquid helium has a nonlinear form for the number distribution function. The function form can be determined using the excitation energy at zero temperature and the latent heat per helium atom at zero temperature. The nonlinear form produces new temperature dependence for the excitation energy from Bose condensate. We evaluate the specific heat using iteration method. The calculation results of the second iteration show good agreement with the experimental data in the temperature region of 0 - 2.1 K, where we have only used the elementary excitation energy at 1.1 K.

\end{abstract}

\section{Nonlinear form of total energy}

Liquid helium system has a total Hamiltonian as
\begin{equation}
H=\sum_{\bf p} {{\frac{\bf{p}^2}{2m}} a^*_{\bf{p}} a_{\bf{p}}} + \frac{1}{2V} \sum_{\bf p , \bf q , \bf k} g({\bf k })a^*_{\bf p +\bf k} a^*_{\bf{q}-\bf k}a_{\bf{p}} a_{\bf{q}}
\end{equation}
where $m$ is the mass of a helium atom, $a^*_{\bf p}$ and $a_{\bf p}$  respectively signify the creation and annihilation operators. We examine the general form of the total energy via the unitary transformation $U$ diagonalizing the Hamiltonian $H$. All eigenstates are described as $|\textrm{ eigenstate}>  = U a^*_{{\bf p}_1} a^*_{{\bf p}_2} a^*_{{\bf p}_3} \cdots a^*_{{\bf p}_N}\textrm{ } |\textrm{ } 0>$ where $\textrm{ } |\textrm{ } 0>$  denotes the vacuum state. New creation and annihilation operators are defined as

\begin{equation}
A^*_{\bf p}=U a^*_{\bf p} U^{-1}, A_{\bf p}=U a_{\bf p} U^{-1}
\end{equation}
which indicate the creation and annihilation operators of a quasi-particle. We designate this quasi-particle as a "dressed boson". The dressed boson number operator is defined as

\begin{equation}
 n_{\bf p} =A^*_{\bf p} A_{\bf p}. 
 \end{equation}
 The total number conservation  and the total momentum conservation are expressed as

\begin{equation}
N=\sum_{\bf q} a^*_{\bf q} a_{\bf q}=\sum_{\bf q} n_{\bf q}
\hspace{1pc}
{\bf Q} =\sum_{\bf q} {\bf q} a^*_{\bf q} a_{\bf q}=U\sum_{\bf q}  {\bf q} a^*_{\bf q} a_{\bf q}U^{-1}=\sum_{\bf q} {\bf q} n_{\bf q} 
\label{eq: a}
\end{equation}
That is to say, the total number of helium atoms is equal to the total number of dressed bosons and the total momentum of helium atoms is equal to the total momentum of dressed bosons.

The total energy of the system is a sum of the kinetic energy $K$ of the center of mass and Galilean invariant terms $X$ : $ K=\frac{{\bf Q}^2}{2M}$  where $M$ is the total mass of liquid helium. 

\begin{eqnarray}
H=\frac{\sum_{\bf p} {\bf p} n_{\bf p} {\cdot} \sum_{\bf q} {\bf q} n_{\bf q} }{2M} +X
= \sum_{\bf p} \frac {{\bf p}^2} {2m} n_{\bf p} - \frac{1}{2M} \sum_{\bf p, q} \frac{1}{2}({\bf p}-{\bf q})^2 n_{\bf p} n_{\bf q} +X
\\= \sum_{\bf p} \frac {{\bf p}^2} {2m} n_{\bf p} +\textrm{(Galilean invariant terms)}
\label{eq: b}
\end{eqnarray}
where Galilean invariant terms are described only by relative momenta of dressed bosons:
\vspace{2pc}

(Galilean invariant terms)=
\begin{equation}
 \frac{1}{N}\sum_{\bf p, q} f_2 ({\bf p}-{\bf q})n_{\bf p} n_{\bf q} +\frac{1}{N^2}\sum_{\bf p, q, k} f_3 ({\bf p}-{\bf q}, {\bf p}-{\bf k})n_{\bf p} n_{\bf q} n_{\bf k}+\dots
\label{eq: c}
\end{equation}\\
Substitution of Eq.(\ref{eq: c}) into Eq.(\ref{eq: b}) yields

 \begin{equation}
E= \sum_{\bf p} \frac {{\bf p}^2} {2m} n_{\bf p} + \frac{1}{N}\sum_{\bf p, q} f ({\bf p}-{\bf q})n_{\bf p} n_{\bf q} 
\label{eq: d}
\end{equation}
where we abbreviate higher terms because three-particle collision is a rare case for diluteness of liquid helium. The single excitation state has a distribution of $\{ n_0=N-1, n_{\bf p} =1 \} $ and therefore its total energy is derived from Eq.(\ref{eq: d}) as follows:

\begin{equation}
E =  f (0)N + \frac {{\bf p}^2} {2m}  + 2( f ({\bf p})-f (0))
\label{eq: e}
\end{equation}
where we have used $1/N \approx 0$ and the spherical symmetric property of the function $f ({\bf p})$. Therein, the latent heat at zero Kelvin is equal to $- N f(0)$. Accordingly the elementary excitation energy at zero Kelvin is given by $\epsilon_p^0 = \frac{{\bf p}^2}{2m} + 2(f({\bf p})-f(0))$. This relation engenders a function form of the nonlinear term as \cite{Sasaki}

\begin{equation}
 f({\bf p})= \frac{1}{2} (\epsilon_p^0 - \frac{{\bf p}^2}{2m} ) + f(0). 
 \label{eq: f}
\end{equation}

\section{Coupled equation determining distribution of dressed bosons}
The energy of one dressed boson is an increase value of the total energy when one dressed boson is added to the system. Accordingly the dressed boson energy is defined as $\omega_{\bf p} = \delta E / \delta n_{\bf p}$. The calculation result for the derivative of Eq. (\ref{eq: d}) shows
\begin{equation}
\omega_{\bf p}(T)= \frac {{\bf p}^2} {2m} + \frac{2}{N}\sum_{\bf q} f ({\bf p}-{\bf q}) n_{\bf q} -  \frac{1}{N^2}\sum_{\bf s, t} f ({\bf s}-{\bf t})n_{\bf s} n_{\bf t} 
\label{eq: g}
\end{equation}
where we have used $ f ({\bf p}-{\bf q}) = f ({\bf q}-{\bf p}) $. The distribution function is determined as
\begin{equation}
n_{\bf p} = \frac {1}{\textrm{exp}((\omega_p (T)-\mu)/(k_B T))-1}
\label{eq: h}
\end{equation}

We can obtain approximate solutions of the coupled equations of (\ref{eq: g}) and (\ref{eq: h}) via the iteration method \cite {Sasaki}. We adopt the Landau distribution function as the zero-th order distribution:
\begin{equation}
n_0 ({\bf p}, T) = \frac {1}{\textrm{exp}(\epsilon_p ^0/(k_B T))-1}
\label{eq: i}
\end{equation}
The j-th order solutions are derived from the (j-1)-th distribution function as follows:
\begin{equation}
\omega_j ({\bf p}, T)= \frac {{\bf p}^2} {2m} + \frac{2}{N}\sum_{\bf q} f ({\bf p}-{\bf q}) n_{j-1} ({\bf q}, T) -  \frac{1}{N^2}\sum_{\bf s, t} f ({\bf s}-{\bf t})n_{j-1} ({\bf s}, T) n_{j-1} ({\bf t}, T) 
\label{eq: j}
\end{equation}
This j-th energy form produces the j-th distribution function:
\begin{equation}
n_j ({\bf p}, T) = \frac {1}{\textrm{exp}((\omega_j ({\bf p}, T)-\omega_j (0, T))/(k_B T))-1}
\label{eq: k}
\end{equation}
Therein the excitation energy from the Bose-Einstein condensate of dressed bosons is expressed as 
\begin{equation}
\epsilon_j ({\bf p}, T) = \omega_j ({\bf p}, T)-\omega_j (0, T) . 
\end{equation}
We can evaluate the second order solutions $ \epsilon_2 ( p, T)$ and $n_2 ( p, T)$ via the iteration processes from the zero-th order distribution.

\section{Evaluation of specific heat } 
Using the second order excitation energy $ \epsilon_2 ( p, T)$ and the distribution function $n_2 ( p, T)$, we can calculate the second order approximation values of specific heat as follows \cite{Sasaki}; 
\\

\begin{figure}[h]
\begin{minipage}{15pc}
\includegraphics[width=15pc]{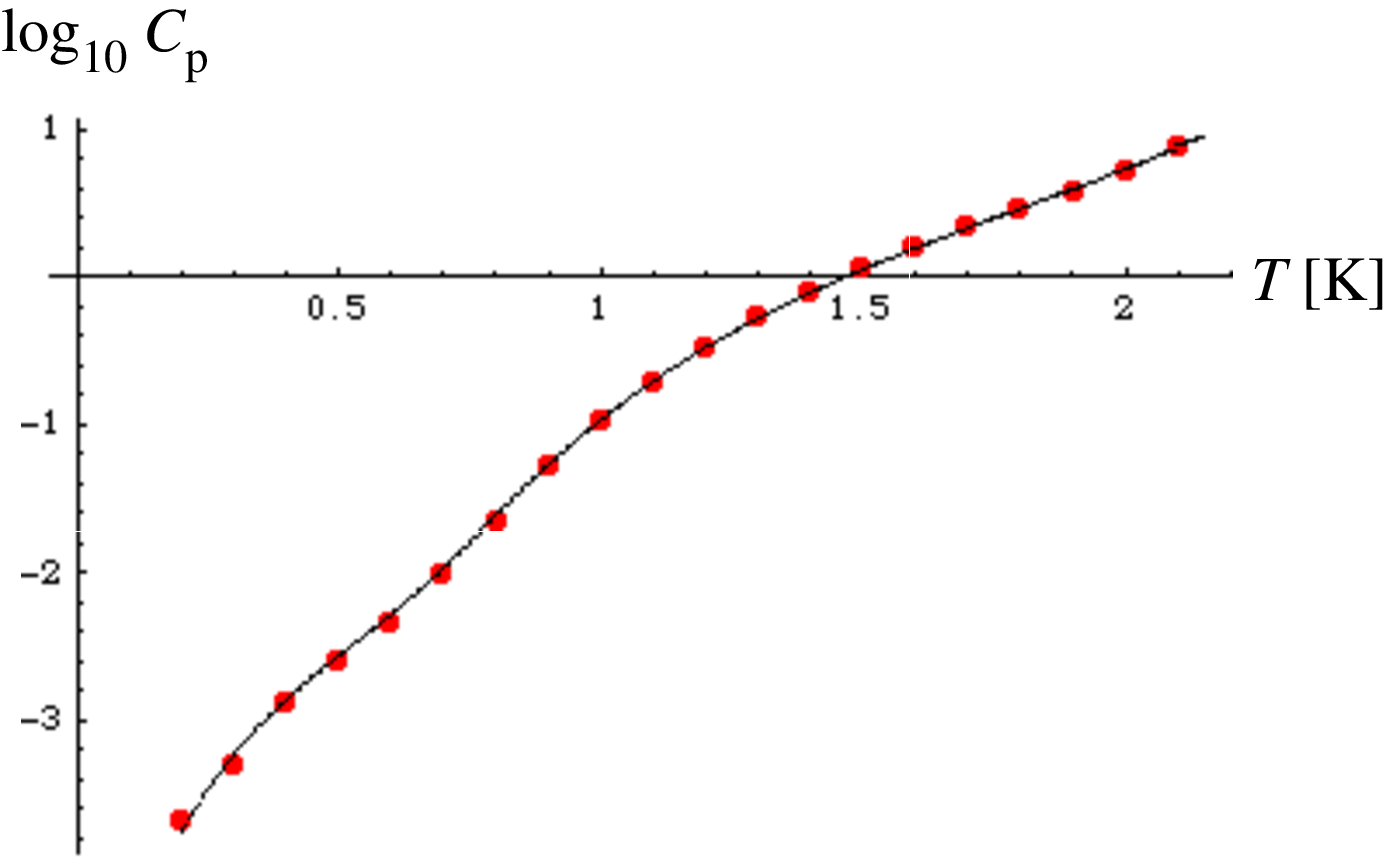}
\caption{\label{f: 1}$C_P$ derived from nonlinear theory. The vertical scale indicates ${\rm log_{10}} C_P$ where $C_P$ is measured by the unit of [J/(g K)]}
\end{minipage}\hspace{1pc}%
\begin{minipage}{14pc}
\includegraphics[width=14pc]{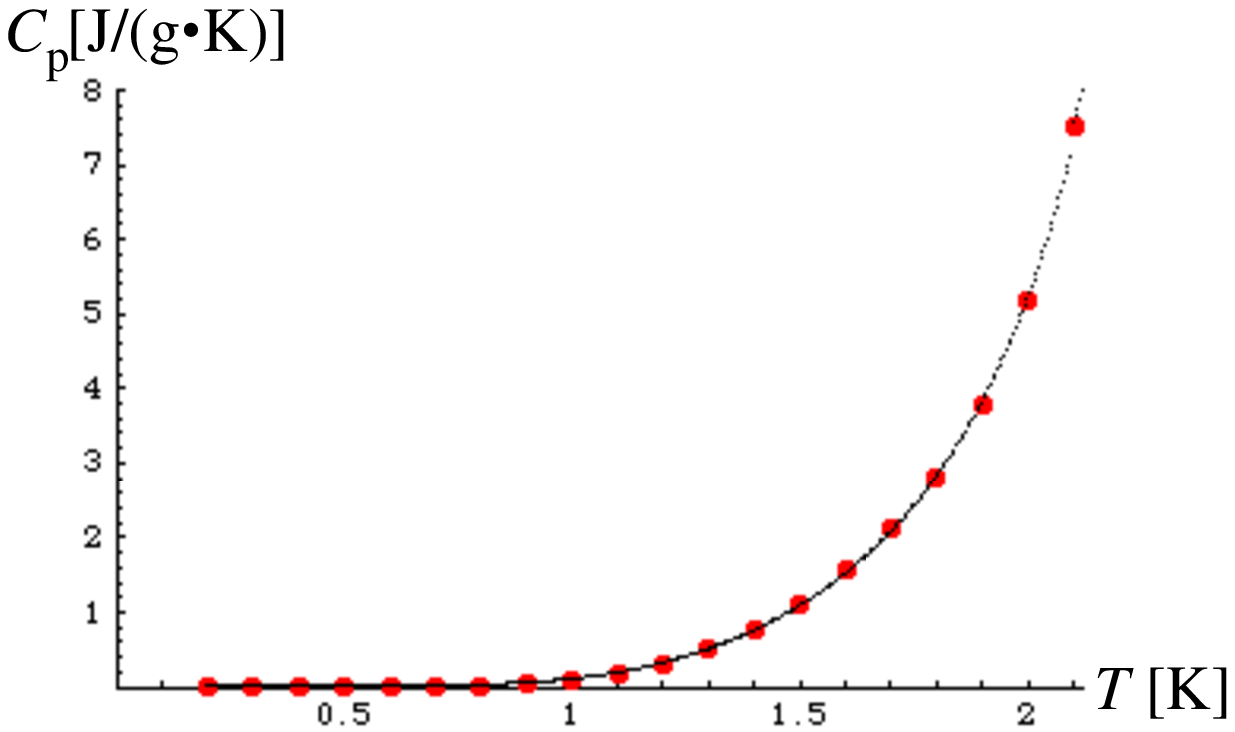}
\caption{\label{f: 2}The curve expresses the calculated values. The dots colored with red indicate experimental data \cite{SpecificHeat}.}
\end{minipage} 
\end{figure}

$C_P = T k_B  \frac {4\pi} {(2\pi \hbar )^3} \Big\{ \frac{\partial V}{\partial T} \Big\}_P \int_0^\infty \Big\{{ \rm log} (1+n_2 (p, T)) +\frac{\epsilon_2 (p, T)}{k_B T} n_2 (p, T) \Big\}p^2 {\rm d}p + 
\\
+  \frac {4\pi V} {(2\pi \hbar )^3} \int_0^\infty (n_2 (p, T))^2 {\rm exp} (\epsilon_2 (p, T)/(k_B T))\times
\\
\times \Bigg\{  \Big\{ \frac{\epsilon_2 (p, T)}{k_B T}   \Big\}^2 k_B  -  \frac{\epsilon_2 (p, T)}{k_B T}   \Big\{ \frac{\partial\epsilon_2 (p, T)}{\partial T}  \Big\}_P  \Bigg\}p^2 {\rm d}p  \hspace{40mm} (17)$
\\\\
The evaluated results are shown in Fig.\ref{f: 1} and Fig.\ref{f: 2}.

\begin{figure}
\begin{center}
\includegraphics[width=33pc]{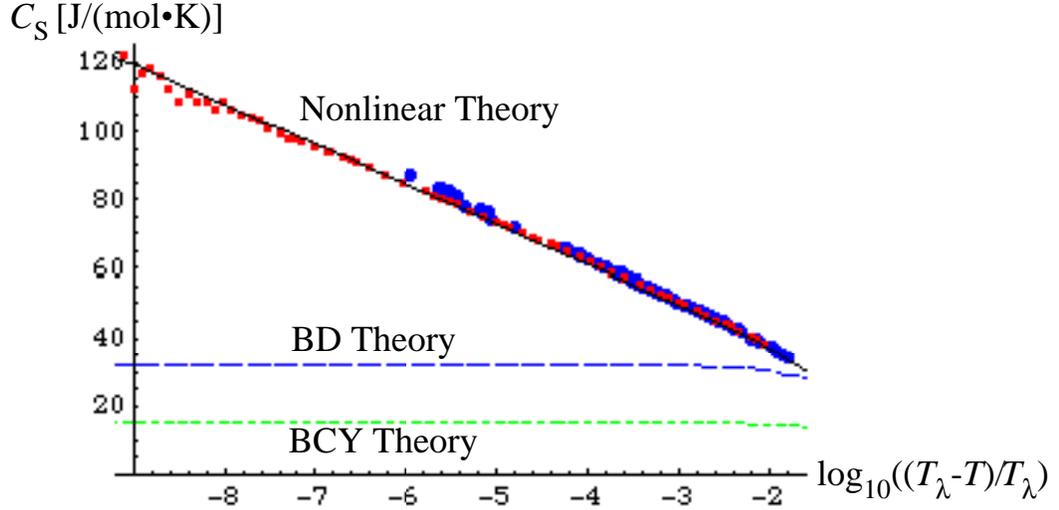}
\end{center}
\caption{\label{f: 3}The large dots colored with blue indicate the experimental data \cite{SpecificHeat}, and the small dots with red are measured by Lipa et al \cite{Lipa}. BD theory indicates the results in the reference of \cite{Brooks}. BCY theory indicates the results in the reference of \cite{Bendt}. The curve of the nonlinear theory is the results of the reference \cite{SasakiLogarithmic}}
\end{figure}

Figure \ref{f: 1} and \ref{f: 2} indicate the second order results of specific heat via the nonlinear theory.
The curves express the calculated values. The dots with red indicate experimental data \cite{SpecificHeat}. 
As shown in Fig.\ref{f: 1} and Fig.\ref{f: 2}, the theoretical values of the second order are in good agreement with the experimental data for $T<2.1  {\rm K}$. It is noteworthy that the present calculation uses the experimental values of excitation energy only for the temperature 1.1 K. Of course the iteration method is insufficient in close vicinity of the $\lambda$ transition temperature. We have discussed origin of the logarithmic divergence at the $\lambda$ point in the previous paper \cite{SasakiLogarithmic}. It is clarified that the logarithmic divergence is caused by the nonlinear mechanism of the total energy. The calculation results are shown in Fig. \ref{f: 3}. 

Thus the nonlinear theory has well explained the temperature dependence of the specific heat of superfluid helium for all temperature region. Accordingly the nonlinear mechanism of total energy is important for understanding the properties of liquid helium.

\end{document}